\documentclass{optica-article}

\journal{opticajournal} 

\articletype{Research Article}

\usepackage{lineno}
\usepackage{xcolor}

\begin{document}

\title{Circular dichroism induction in WS$_2$ by a chiral plasmonic metasurface}

\author{Fernando Lor\'en,\authormark{1,2} Cyriaque Genet,\authormark{3} and Luis Martin-Moreno\authormark{1,2,*}}
  
\address{\authormark{1}Instituto de Nanociencia y Materiales de Arag\'on (INMA), CSIC-Universidad de Zaragoza, 50009 Zaragoza\\
\authormark{2}Departamento de F\'isica de la Materia Condensada, Universidad de Zaragoza, 50009 Zaragoza\\
\authormark{3}University of Strasbourg and CNRS, CESQ \& ISIS (UMR 7006), 8, all\'ee G. Monge, 67000 Strasbourg, France}

\email{\authormark{*}lmm@unizar.es} 



\begin{abstract*}
We investigate the interaction between a monolayer of  WS$_2$ and a chiral plasmonic metasurface.  WS$_2$ possesses valley excitons that selectively couple with one-handed circularly polarized light. At the same time, the chiral plasmonic metasurface exhibits spin-momentum locking, leading to a robust polarization response in the far field. Using a scattering formalism based on the coupled mode method, we analyze various optical properties of the WS$_2$ monolayer. Specifically, we demonstrate the generation of circular dichroism in the transition metal dichalcogenide (TMD) by harnessing the excitation of surface plasmon polaritons (SPPs) in the metasurface. Moreover, we observe the emergence of other guided modes, opening up exciting possibilities for further exploration in TMD-based devices.
\end{abstract*}

\section{Introduction}
Transition metal dichalcogenides (TMDs), such as MoS$_2$, WS$_2$, and MoSe$_2$, have garnered significant attention due to their unique electronic, optical, and mechanical properties \cite{wang2012electronics, baugher2014optoelectronic, zhang2015large, eginligil2015dichroic, manzeli20172dtransition,  li2019valley, li2020roomtemperature}. TMDs consist of atomically thin layers held together by weak van der Waals forces. The monolayers of these materials exhibit fascinating characteristics that make them highly desirable for a range of applications \cite{mak2010atomically, bie2017mote2based}, including electronics \cite{mak2014thevalley}, optoelectronics \cite{yang2016electrically, wan2017epitaxial}, energy storage \cite{bernardi2013extraordinary} and sensing \cite{lee2018twodimensional}. Furthermore, circularly polarized light is particularly relevant in this context, as it enables the selective excitation and manipulation of the characteristic valley excitons of the TMDs \cite{schaibley2016valleytronics, mccreary2017understanding, gong2018nanoscale}. 

On the other hand, plasmonic metasurfaces have demonstrated a wide range of applications \cite{chen2016areview, rodrigo2016extraordinary, genevet2017recent} such as sensing \cite{beruete2019terahertz} or imaging \cite{watts2014terahertz, walter2017ultrathin}. Chiral structures, in particular, have emerged due to their capacity to manipulate and control the polarization of light waves \cite{bomzon2002spacevariant, zhao2011manipulating, yu2012abroadband, shitrit2013spinoptical, langguth2015plasmonic, cotrufo2016spindependent, yan2017twisting}. Furthermore, they can exhibit spin-momentum locking (SML), which refers to the coupling between the polarization and momentum of the light waves involved \cite{bliokh2015spinorbit}. The SML of the metasurface enables a particularly compelling application in the field of valleytronics \cite{chervy2018room, li2018tailoring, jha2018spontaneous, guddala2019valley, sun2019separation, rong2020photonic}, where the spin angular momentum of emitted light can be used to excite and detect valley excitons selectively.

Significant advancements have been reported in recent years for achieving enhanced valley polarization or circular dichroism (CD) in TMD materials through the application of electrical, magnetic, and optical bias \cite{kim2020asingle, cao2012valleyselective, lin2021electrically, guddala2021optical, li2016tailoring, guddala2021alloptical}. These developments primarily involve encapsulated TMD monolayers or the integration of TMD materials with plasmonic or dielectric metasurfaces and waveguides. In this work, as proof of concept, we study the system composed of a WS$_2$ monolayer over a chiral plasmonic metasurface composed of rotated dimples to analyze the CD induced in the WS$_2$. Finally, we show that guided modes can enhance the CD in the presence of a dielectric spacer between the TMD and the metasurface. Unlike SPPs, these modes have a transverse electric (TE) field, but we observe that mutatis mutandis, its excitation satisfies the 3-step model derived in \cite{loren2023microscopic}.

\section{Theoretical description}
\begin{figure}[ht!]
\centering\includegraphics[width=0.7\textwidth]{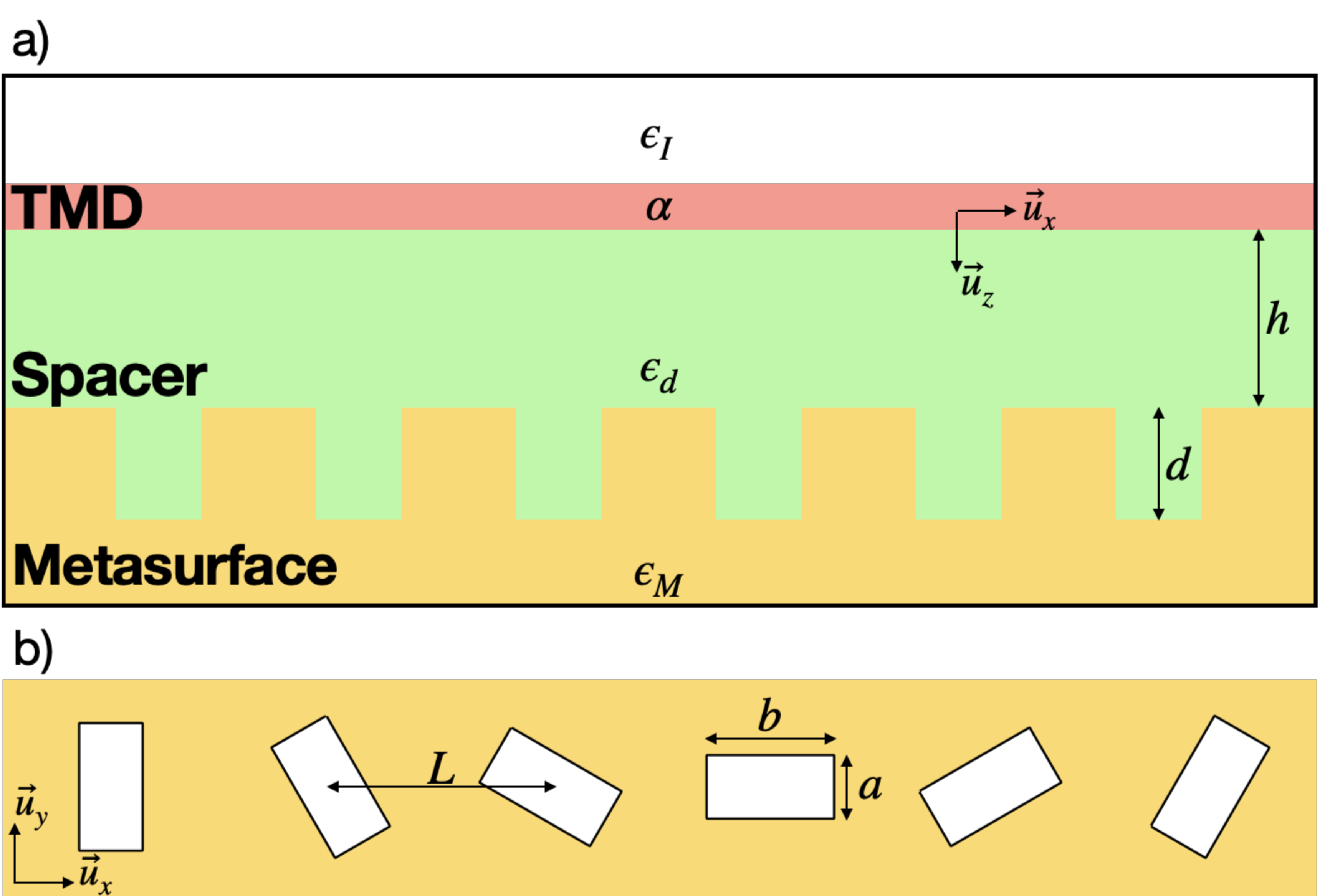}
\caption{a) Scheme of the system studied in the article, with the corresponding defining magnitudes. b) Unit cell with $N=6$ and $n_w = 1/2$. Based on \cite{chervy2018room}.}
\label{fig:scheme}
\end{figure}

We consider a WS$_2$ monolayer deposited over a chiral plasmonic metasurface, with a dielectric spacer in between (see Figure~\ref{fig:scheme}(a)), which is based on \cite{chervy2018room}.  As in the experiments in Ref. \cite{chervy2018room}, the metasurface is characterized by a periodic array of $N=6$ rectangular dimples step-wisely rotated with a winding number of $n_w = 1/2$, which is the number of complete $2\pi$ rotations along the whole unit cell. The unit cell (see Fig.~\ref{fig:scheme}(b)) is periodically repeated in both $\vec{u}_x$ and $\vec{u}_y$ directions.  

We use the coupled-mode method (CMM) to derive the system's electromagnetic (EM) fields. This method has been extensively employed to study EM properties in plasmonic arrays \cite{lmm2008minimal, deleon2008theory, garciavidal2010light, loren2023microscopic}. The CMM expands the EM fields in plane waves in the free space regions and waveguide modes inside the dimples and finds the electric field amplitudes by adequately matching the EM fields at the interfaces. The excitation of our system consists of an electromagnetic plane wave impinging our metasurface with an in-plane wavevector $\vec{k}^{in} = k_x^{in} \vec{u}_x + k_y^{in} \vec{u}_y$ and an incident polarization $\sigma^{in}$. The goal is to compute the coefficients associated with the Bragg orders, which define the EM fields in the different regions. 

We do not present the full equations resulting from the CMM method, as they have been given in a similar structure without the TMD in \cite{loren2023microscopic}. The effect of the 2D material is straightforwardly included by considering an additional interface and satisfying the in-plane magnetic field discontinuity that arises due to the conductivity $\alpha$. 

The accuracy of the CMM relies on the number of considered Bragg modes in the simulations. In our particular case, we have considered $21$ modes in the $\vec{u}_x$ direction and $3$ modes in the $\vec{u}_y$ direction, being $\vec{G}_x = 2 \pi / (6 L) \, \vec{u}_x$ and $\vec{G}_y = 2 \pi / L \, \vec{u}_y$ the reciprocal lattice or Bragg vectors of the unit cell \cite{fox2022generalized}.  We have checked that adding more Bragg modes introduces less than $1\%$ variation in the obtained results.

We treat the metallic structure using the surface impedance boundary conditions (SIBC) approximation, which considers the actual dielectric constant of the metal $\epsilon_M$ (via the Lorentz-Drude model \cite{vial2005improved}) and the penetration of the EM fields into the metal slab through the surface impedance $z_s = 1/\sqrt{\epsilon_M + 1}$, which leads to the exact dispersion relation of SPPs in a metal-vacuum interface. The semi-infinite region above the 2D material and the dielectric spacer is characterized by the dielectric constants $\epsilon_I$ and $\epsilon_d$, respectively. Finally, the 2D material is characterized by its 2D conductivity $\alpha$.  We obtain it from the experimental dielectric constant $\epsilon_{2D}$ \cite{liu2014optical} through $\alpha = w \pi (\epsilon_{2D}-1)/(\lambda i)$, where $w$ is the width of the 2D material ($w\simeq 1 \, nm$), $\lambda$ is the wavelength of the EM fields and Gaussian units have been used. Moreover, we fit the conductivity to remove the background, which is associated with sample imperfections and can be decreased by, for instance, encapsulating the TMD between graphene monolayers \cite{lorchat2018roomtemperature, lorchat2020filtering}. 

Once the EM fields are computed, we obtain the total absorptance of the system as $A_T = (W_{inc} - W_{ref})/W_{inc}$, where $W_{inc}$ and $W_{ref}$ are the incoming and reflected EM energy fluxes, respectively, calculated via the integration of the Poynting vector over an $x-y$ plane in the semi-infinite regions above the 2D material. The reflectance is $R = W_{ref} / W_{inc} = 1 - A_T$. The absorptance of the WS$_2$ monolayer depends on the conductivity tensor $\overleftrightarrow{\alpha}$ and the in-plane electric field $\vec{E}$ at the 2D material, integrated over the $x-y$ plane: $A_{WS_2} = \left(\int \vec{j}^* \vec{E} \, dx \, dy\right) / W_{inc} = \left( \int \vec{E}^* \overleftrightarrow{\alpha} \vec{E} \, dx \,dy \right) / W_{inc}$. In the linear polarization basis $\overleftrightarrow{\alpha}$ is diagonal, with diagonal components equal to $\alpha$.

Associated with them, we also analyze the circular dichroism and the g-factor: $CD = A_{WS_2}^+ - A_{WS_2}^-$, $CD_{norm} = CD / (A_{WS_2}^+ + A_{WS_2}^-)$, $g = R^+ - R^-$, and $g_{norm} = g/(R^+ + R^-)$. The superscripts refer to the spin of the incoming plane waves $\sigma^{in} = \pm$, and we will work with both normalized and not-normalized magnitudes. 

For definiteness, the geometrical parameters considered in this paper are as in Ref.  \cite{chervy2018room}: each dimple has a short side $a=80\,nm$, a long side $b=220\,nm$ and a depth $d=60\,nm$, and the distance between the centers of nearest dimples is $L=480\,nm$ (in both $\vec{u}_x$ and $\vec{u}_y$).

\section{Circular Dichroism in WS$_2$}
To analyze the quantities described in the previous section, we range both the energy $\omega$ and the momentum $k_x^{in}$ of the incoming plane wave. 

\begin{figure}[ht!]
\centering\includegraphics[width=\textwidth]{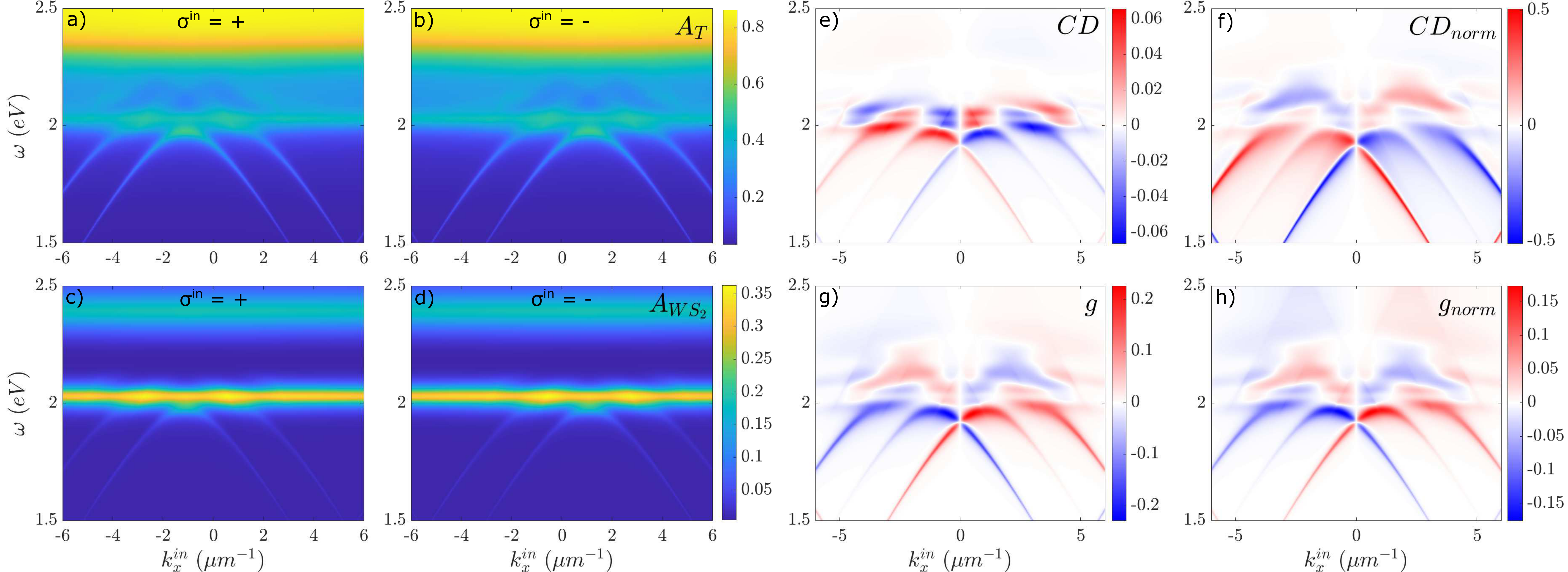}
\caption{(a,b) Total absorptance of the system for $\sigma^{in} = \pm$, respectively. (c,d) Absorptance of the WS$_2$ monolayer for $\sigma^{in}=\pm$, respectively. (e) Circular dichroism. (f) Normalized circular dichroism. (g) g-factor. (h) Normalized g-factor. For all the plots, we have considered a distance $h=10\,nm$ between the metasurface and the WS$_2$, above the 2D material $\epsilon_I = 1$ (vacuum), and a spacer dielectric constant $\epsilon_d = 2.5$, which corresponds to the typical $SiO_2$ spacer.}
\label{fig:CDinduction}
\end{figure}

In Figs.~\ref{fig:CDinduction}(a-d), we observe the well-known behavior of both the metasurface and the TMD monolayer. In the upper region of the $A_T^{\pm}$ figures, we identify the large absorption of gold at $\lambda \sim 500 \, nm$. Moreover, the four absorptance figures present plasmonic parabolic features provided by the metasurface. The SML, produced by the rotation of the dimples along the unit cell, is also noted through the selective excitation of the left and right SPPs (which depend on the spin $\sigma^{in} = \pm$ of the incoming plane wave, respectively \cite{chervy2018room, loren2023microscopic}).  The influence of the WS$_2$ monolayer is also seen in all absorptance figures. Both exciton bands are apparent:  The A-exciton band appears at $\omega \simeq 2.03 \, eV$, and the B-exciton band at $\omega \simeq 2.4 \, eV$. Besides, in Figs.~\ref{fig:CDinduction}(c,d), we observe the interplay between the A-exciton band from the WS$_2$ monolayer and the SPPs through the absorptance enhancement at the band's crossing points.  

The same metasurface was previously studied in \cite{loren2023microscopic}, showing that \textit{spin-momentum locking} is an approximate symmetry.  The symmetry breakdown arises because the EM plane waves are transversal with respect to their momentum, while the interaction with the surface presents symmetries with respect to the metasurface normal.  Consequently, when circularly polarized light is projected onto the planar surface, it transforms into an elliptical polarization state comprising a combination of $\pm$ spin states, spoiling the SML effect. This is not apparent in the contour plots in Fig.~\ref{fig:CDinduction} because of their limited resolution. However, if we cut Fig.~\ref{fig:CDinduction}(a) at fixed energy, we would observe two small absorptance peaks at the $k_x^{in}$ for which plasmon with the opposite spin states appear.

Figs.~\ref{fig:CDinduction}(e,f) show the computed CD and the $CD_{norm}$. We observe a different absorptance in the WS$_2$ depending on the incident spin $\sigma^{in}$.  Notice that, without the chiral metasurface,  the $K$ valley would preferentially absorb $+$ light (in fact, it would only absorb $+$ light at normal incidence) and the same for the $K'$ valley with respect to $+$ light, but this would lead to $CD=0$. The non-zero CD in the TMD monolayer induced by the chiral plasmonic array is maximal when the spin-momentum-locked SPPs are excited. In Fig.~\ref{fig:CDinduction}(e), around the A-exciton band, we observe a difference in absorptances of $\sim0.06$, which is significant considering that the typical absorptance in that band is $\sim 0.3$. Besides, in Fig.~\ref{fig:CDinduction}(f), the plasmonic parabolas present $|CD_{norm}| \simeq 0.5$, which implies that the WS$_2$ absorptance for one polarization is three times larger than for the opposite.

In Figs.~\ref{fig:CDinduction}(g,h), we represent $g$ and $g_{norm}$. Their behavior is inverse to those of CD or $A_T^{\pm}$ because they are associated with the reflectance $R^{\pm}$. For example, $g_{norm}$ at the rightmost parabola is positive because the chiral metasurface reflects the $+$ light while  $-$ light excites an SPP and is thus partially absorbed.  In general, $g_{norm}$ is mainly determined by the chiral plasmonic metasurface, not being strongly affected by the presence of the TMD. 

These results are just proof of how a chiral plasmonic array can induce CD in a 2D material. The circular dichroism or other quantities could be optimized by changing geometrical parameters such as the distance between the dimples, their size and depth, the distance between the metasurface and the 2D material, or the dielectric constant of the spacer, among others.


\section{Influence of the TE waveguide modes}
The second main result of this article is developed in this section, where we show that the spacer can induce the existence of a set of guided modes with imprinted chiral properties from the chiral holey surface.  

\begin{figure}[ht!]
\centering\includegraphics[width=\textwidth]{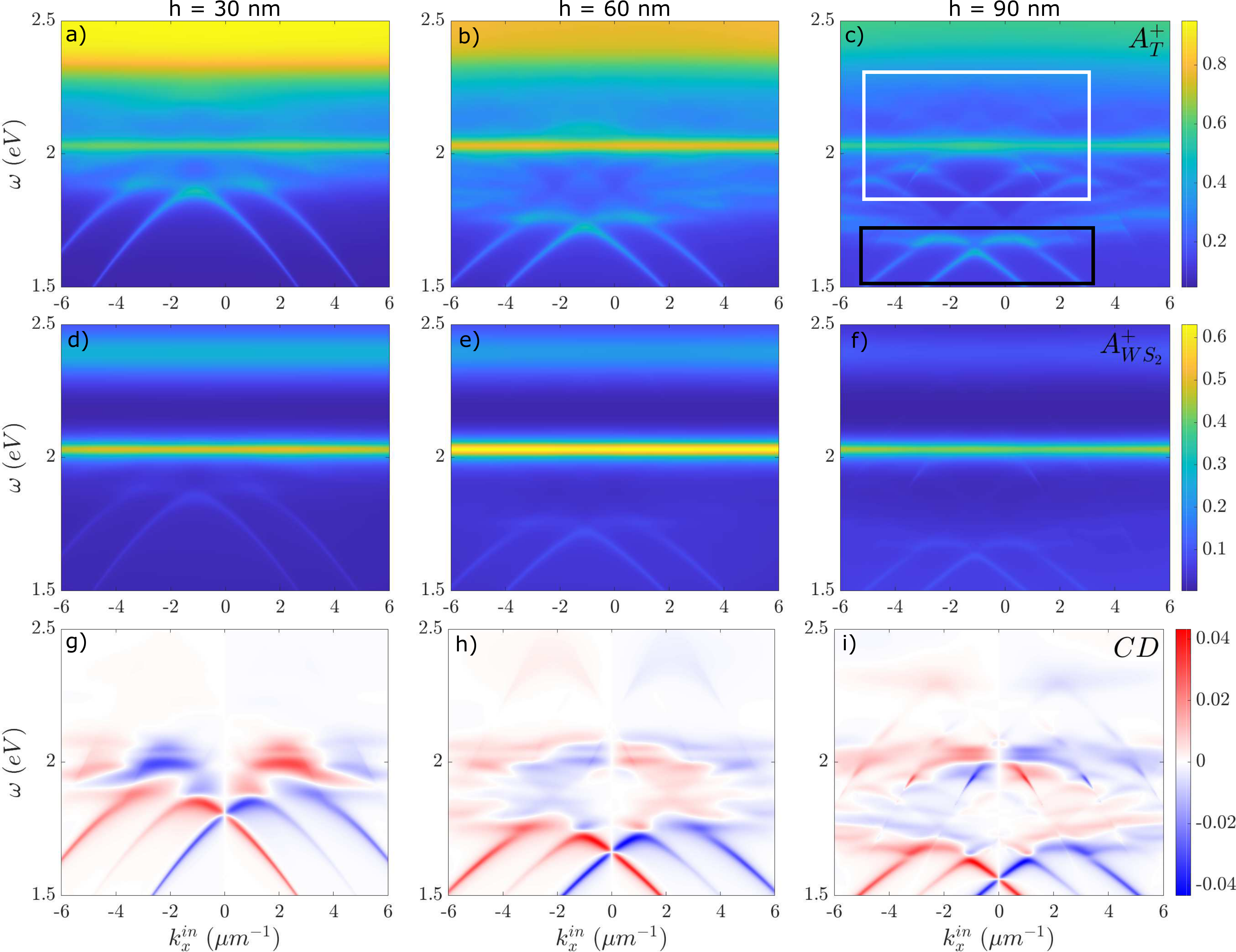}
\caption{(a,b,c) Total absorptance, (d,e,f) WS$_2$ monolayer absorptance, and (g,h,i) circular dichroism, considering $\epsilon_I = 1$, $\epsilon_d = 2.5$ and $\sigma^{in} = +$. For different distances from the metasurface to the WS$_2$ monolayer: (a,d,g) $h = 30 \, nm$, (b,e,h) $ h = 60 \, nm$, and (c,f,i) $h = 90 \, nm$. The white (black) box highlights the guided (plasmonic) modes.}
\label{fig:Abs_hs}
\end{figure}

In Figs.~\ref{fig:Abs_hs}(a-c), we represent $A_T^+$ for three different distances between the metasurface and the WS$_2$ monolayer, $h$. The SPPs parabolic branches are excited at lower incident energies as $h$ increases.  At the same time, we observe the appearance of new parabolas for larger $h$. These are the emerging guided modes (white box). Besides, in Figs.~\ref{fig:Abs_hs}(d-i), we represent $A_{WS_2}^+$ and CD for the three different $h$'s to prove that a larger absorption in the monolayer does not imply a larger CD in it.

\begin{figure}[ht!]
\centering\includegraphics[width=\textwidth]{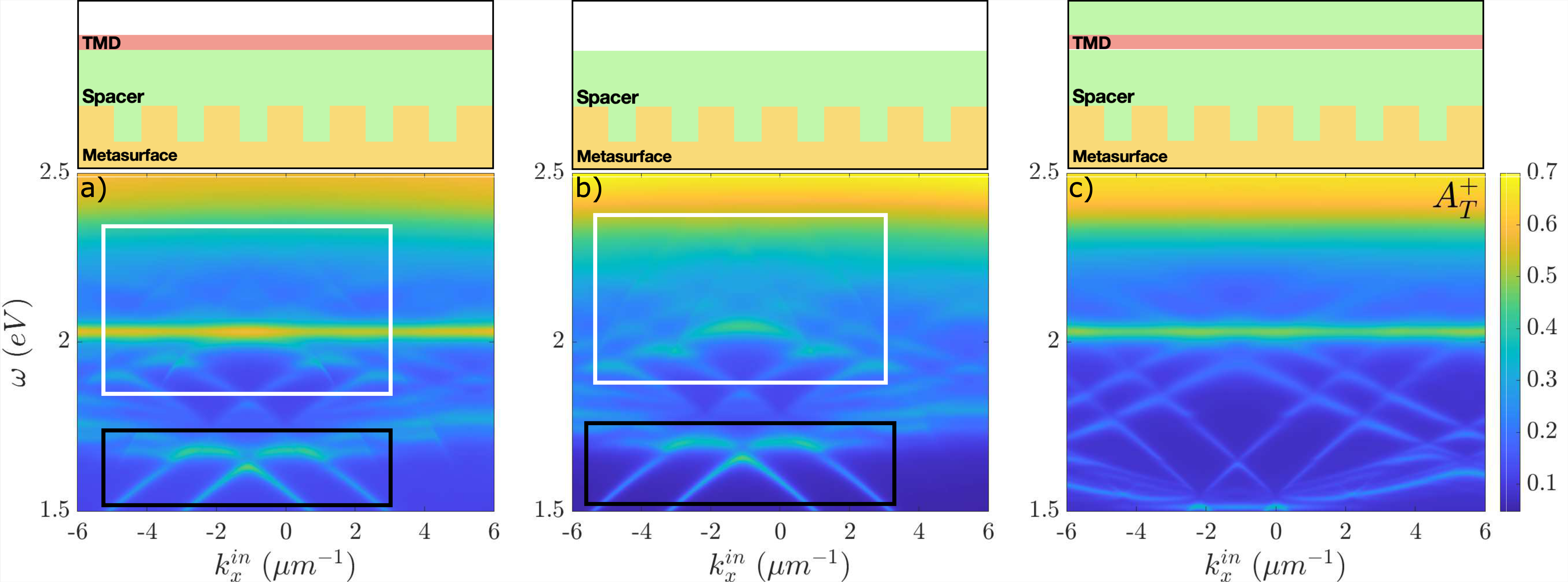}
\caption{Total absorptance, considering $\epsilon_I = 1$, $\epsilon_d = 2.5$, $h=90\,nm$ and $\sigma^{in} = +$. For the different structures above each contour plot: (a) With the WS$_2$ monolayer (same as Fig.~\ref{fig:Abs_hs}(c)), (b) Without the WS$_2$ monolayer, and (c) With the WS$_2$ monolayer and $\epsilon_I = 2.5$. The white (black) boxes highlight the guided (plasmonic) modes.}
\label{fig:Abs_noWS2}
\end{figure}

To understand the origin of these guided modes, Fig.~\ref{fig:Abs_noWS2} renders $A_T^+$ for three different situations, sketched above the plots. Fig.~\ref{fig:Abs_noWS2}(a) is the same as Fig.~\ref{fig:Abs_hs}(c). Fig.~\ref{fig:Abs_noWS2}(b) is computed by removing the WS$_2$ from the top of the dielectric spacer. Fig.~\ref{fig:Abs_noWS2}(c) considers the TMD, but with the region above the WS$_2$ also having dielectric constant $\epsilon_I = 2.5$. These modes already appear in Fig.~\ref{fig:Abs_noWS2}(b) when no WS$_2$ monolayer is present. Besides, only the SPPs are excited if the semi-infinite superstrate is considered to have the same dielectric constant of the spacer, even if we consider the presence of the WS$_2$ layer. Thus, the guided modes are due to the presence of the spacer and not to the TMD (although the 2D material slightly affects the required energy for their excitation). 

\begin{figure}[ht!]
\centering\includegraphics[width=\textwidth]{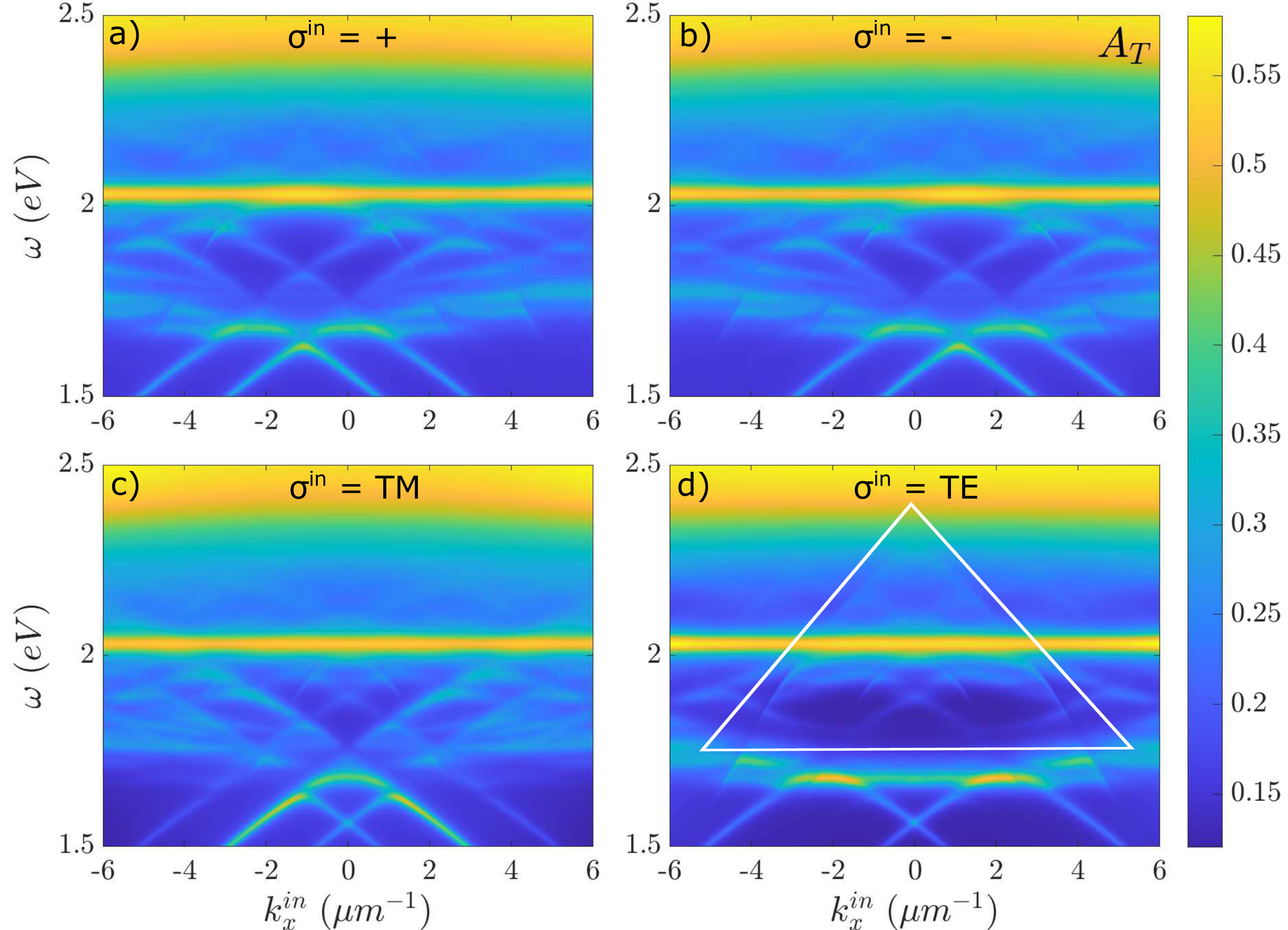}
\caption{Total absorptance, considering $h=90\,nm$ and $\epsilon_d = 2.5$. For incident polarization: (a) $\sigma^{in} = +$, (b) $\sigma^{in} = -$, (c) $\sigma^{in} = p \equiv TM$, and (d) $\sigma^{in} = s \equiv TE$. The triangular white box highlights the guided TE-guided mode parabola.}
\label{fig:Abs_TEmodes}
\end{figure}

These spacer-induced guided modes are TE-polarized. To show how this changes the polarization response of the considered structure, we represent in Fig.~\ref{fig:Abs_TEmodes} the total absorptance of the system $A_T$ for four incident polarizations. Figs.~\ref{fig:Abs_TEmodes}(a,b) also show that the SML holds for the guided modes. However, for linearly polarized light with $\sigma^{in} = TM$ (see Fig.~\ref{fig:Abs_TEmodes}(c)), absorptance is enhanced at the central plasmonic parabola but vanishes for the guided mode, agreeing with the known TM character of SPPs. Complementarily, for $\sigma^{in} = TE$ (see Fig.~\ref{fig:Abs_TEmodes}(d)), we observe just the opposite: an enhancement for the central parabola of the guided mode and a substantial reduction for the corresponding SPP. Using the analogy of the SPPs, this confirms the TE character of these guided modes.

\begin{figure}[ht!]
\centering\includegraphics[width=\textwidth]{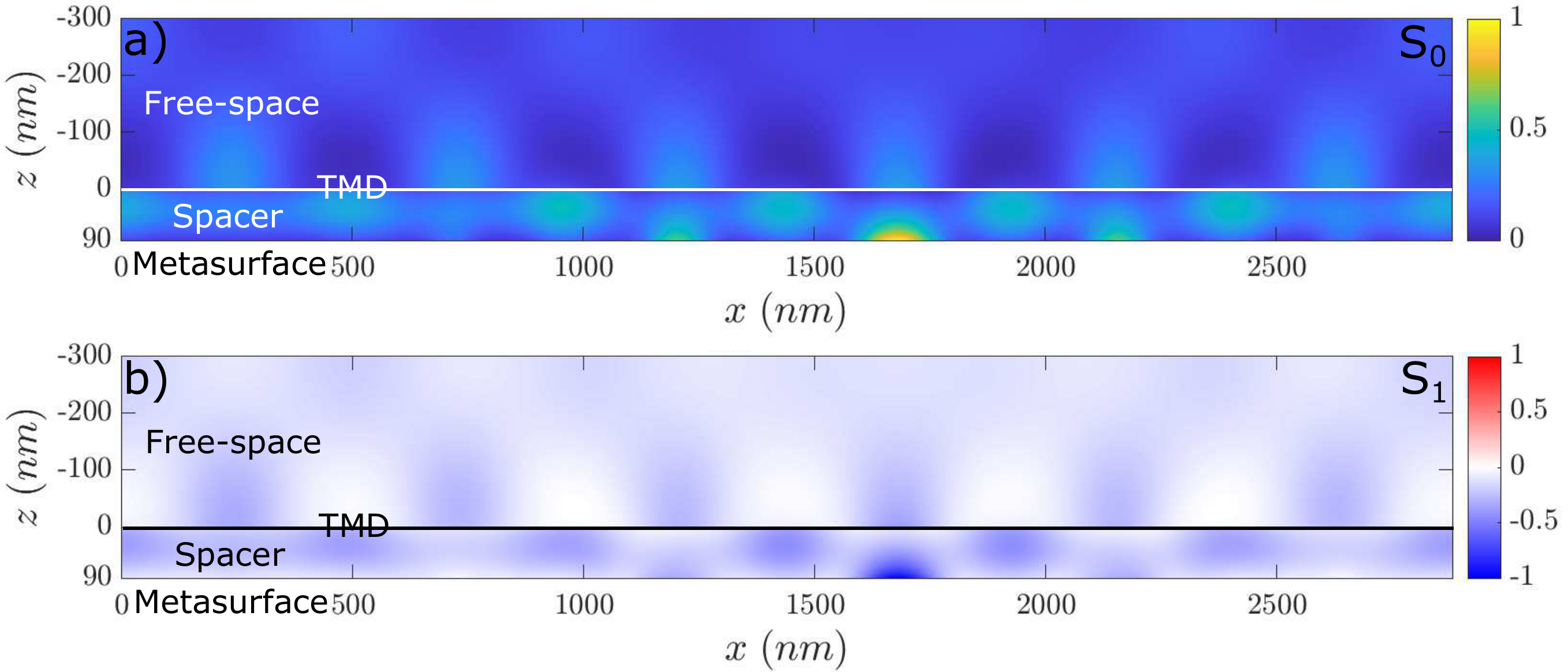}
\caption{Field distributions: (a) Zeroth component of the Stokes vector $S_0$, and (b) First component of the Stokes vector $S_1$; considering $h=90\,nm$, $\epsilon_d = 2.5$, incident polarization $\sigma^{in} = s \equiv TE$, incident energy $\omega = 2.28 \, eV$ and impinging normally to the metasurface. Both components are normalized with respect to the maximum value of $S_0$.}
\label{fig:fields}
\end{figure}

Through Figs.~\ref{fig:Abs_noWS2} and \ref{fig:Abs_TEmodes} we have shown that the guided modes are confined and TE. However, a visual representation of the field and polarization profiles can prove this point beyond any doubt. For this, we have computed several field distributions in Fig.~\ref{fig:fields}. We have decided to represent the zeroth (see Fig.~\ref{fig:fields}(a)) and first (see Fig.~\ref{fig:fields}(b)) components of the Stokes vector in real space: $S_0$ and $S_1$, respectively, which encapsulates the total amplitude and difference between the TM and TE polarizations for the in-plane electric field. This is: $S_0 = |E_x|^2 + |E_y|^2$ and $S_1 = |E_x|^2 - |E_y|^2$, which indicates whether the electric field is transverse ($S_1<0$) or longitudinal ($S_1>0$), considering that the resonant in-plane momentum component points along $\vec{u}_x$. Therefore, a larger $S_0$ and a negative $S_1$ distribution in the spacer region would confirm the confinement and that the electric field generated by the resonance is TE, respectively. In Fig.~\ref{fig:fields}, we represent the $z-x$ plane, fixed at $y = 240\, nm$ (middle of dimples) and impinging normally to our system with a plane wave with energy $2.28 \,eV$ and TE polarization, i.e., just in the guided mode resonance. $x$ covers the unit cell length, whereas $z$ goes from $z = 90\, nm$ (metal surface) to $z = -300 \,nm$ (free-space region), crossing $z = 0\, nm$ where the WS$_2$ monolayer is placed. From $S_0$ is seen that the modes are confined, whereas, from $S_1$, one can check that the light is mostly TE.

Note that the modified SML rules, developed in \cite{loren2023microscopic} to address the resonant excitation of plasmons, also apply to these guided modes when slightly modified to incorporate their TE nature properly.  Scattering under resonant excitation can be understood as a three-step scattering process. First, the incoming plane wave picks up grating momentum to excite the SPP/guided mode, ending with polarization dictated by the SML rules. In the second step, this polarization is projected onto TM/TE polarization states due to the linear polarization of the confined modes. Lastly, the SPP/guided mode outcouples via another SML process. Both in- and out-coupling processes can be ``normal'' (maintaining the polarization) or ``chiral'' (which reduces or increases the plane wave spin when the geometric momentum $k_g$, associated with the rotation of the dimples, is added or subtracted).

\begin{figure}[ht!]
\centering\includegraphics[width=\textwidth]{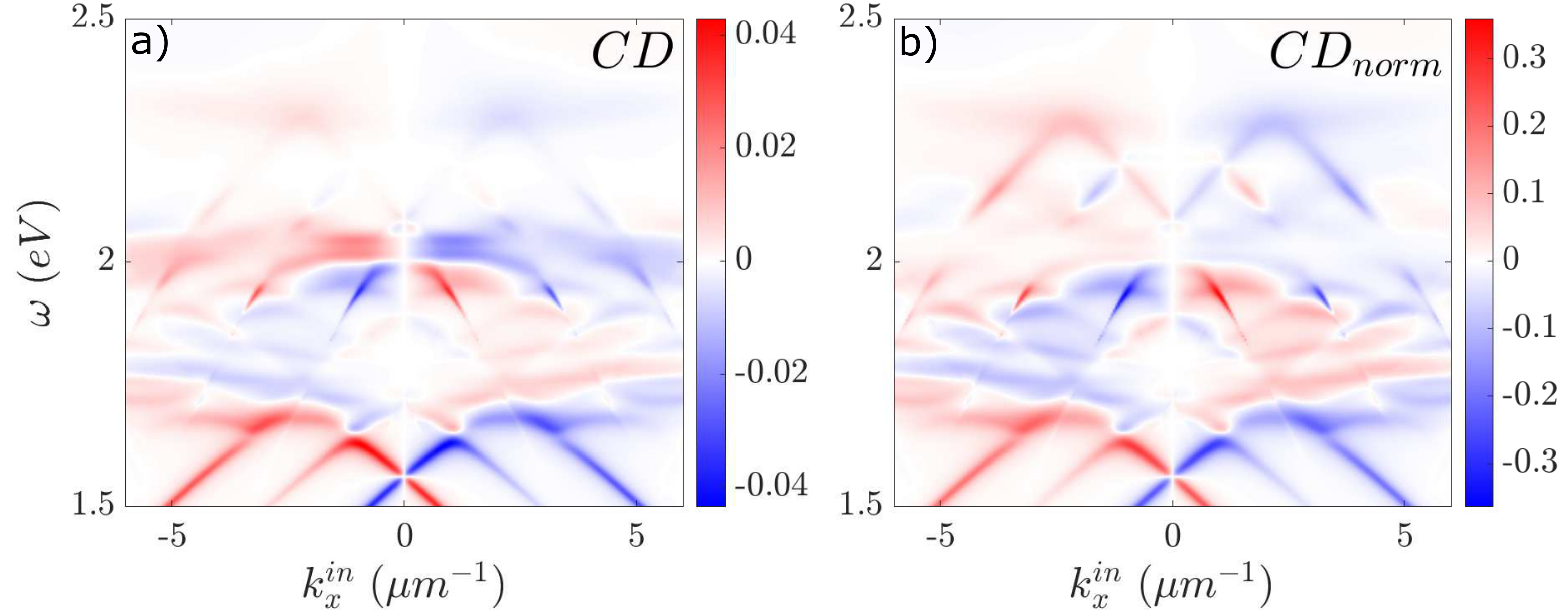}
\caption{(a) Circular dichroism. (b) Normalized circular dichroism. We have considered $h=90\,nm$ and $\epsilon_d = 2.5$.}
\label{fig:CD_TEmodes}
\end{figure}

To conclude, in Fig.~\ref{fig:CD_TEmodes}, we represent the CD and $CD_{norm}$ considering $h=90\,nm$, $\epsilon_I = 1$ and $\epsilon_d = 2.5$. The TE-guided modes also induce circular dichroism in the WS$_2$ monolayer. We notice $|CD|\simeq0.04$ around the A-exciton band and $|CD_{norm}|\simeq0.3$ when we excite one of these modes. These magnitudes are not as large as those Figs.~\ref{fig:CDinduction}(e,f) but demonstrate inducing chiral behavior in the WS$_2$ monolayer through the TE-guided modes, and the possibility to use the spacer layer to enhance this phenomenon further.

\section{Conclusion}
We have presented a way to obtain circular dichroism in transition metal dichalcogenides using chiral plasmonic metasurfaces. Additionally, a transverse electric waveguide mode has been observed and demonstrated to cause CD in TMDs. This work is a proof of concept in a non-optimized structure; CD values are expected to be substantially higher after optimization over the whole range of parameters.  In particular,  after optimization, the transverse electric character of the guided modes is expected to couple them better than SPPs to the TMDs. 

Our characterization of the 2D material is simple, only based on its permittivity tensor. By changing this tensor into the circularly polarized basis, we can easily split the full TMD optical behavior into the $K$ and $K'$ valleys components. This allows us to study the different valley phenomena with our simple formalism. However, magnitudes such as the valley contrast or the valley-selective circular dichroism would still require considering the inter-valley scattering to embrace the exciton dynamics, which has been proved to be essential in the analysis of TMDs \cite{chen2017valleypolarized, dufferwiel2018valley, chervy2018room}. 

The findings presented here pave the way for further exploration in the field of chiral optics and valleytronics. Indeed, the induced circular dichroism in TMDs may have applications in areas such as chiral sensing, imaging, and quantum information processing \cite{schaibley2016valleytronics, schneider2018twodimensional, li2019valley, sun2020selectively, zheng2023electroninduced, gao2023atomicallythin}.

\begin{backmatter}
\bmsection{Funding}
We acknowledge Project PID2020-115221GB-C41 was financed by MCIN/AEI/10.13039/501100011033.

This work is part of the Interdisciplinary Thematic Institute QMat of the University of Strasbourg, CNRS, and Inserm. It was supported by the following programs: IdEx Unistra (ANR-10-IDEX-0002), SFRI STRATUS project (ANR-20-SFRI-0012), and USIAS (ANR-10-IDEX-0002-02), under the framework of the French Investments for the Future Program.

\bmsection{Acknowledgments}
FL and LMM also acknowledge the Aragón government project Q-MAD.


\bmsection{Disclosures}
The authors have no conflicts to disclose.

\bmsection{Data Availability Statement}
Data underlying the results presented in this paper are not publicly available but may be obtained from the authors upon reasonable request.

\end{backmatter}


\bibliography{draft}

\begin{thebibliography}{10}
\newcommand{\enquote}[1]{``#1''}

\bibitem{wang2012electronics}
Q.~H. Wang, K.~Kalantar-Zadeh, A.~Kis, J.~N. Coleman, and M.~S. Strano,
  \enquote{Electronics and optoelectronics of two-dimensional transition metal
  dichalcogenides,} {\protect\JournalTitle{Nature Nanotechnology}} \textbf{7},
  699--712 (2012).

\bibitem{baugher2014optoelectronic}
B.~W.~H. Baugher, H.~O.~H. Churchill, Y.~Yang, and P.~Jarillo-Herrero,
  \enquote{Optoelectronic devices based on electrically tunable p{\textendash}n
  diodes in a monolayer dichalcogenide,} {\protect\JournalTitle{Nature
  Nanotechnology}} \textbf{9}, 262--267 (2014).

\bibitem{zhang2015large}
Q.~Zhang, S.~A. Yang, W.~Mi, Y.~Cheng, and U.~Schwingenschlögl, \enquote{Large
  spin-valley polarization in monolayer {MoTe}$2$ on top of {EuO}(111),}
  {\protect\JournalTitle{Advanced Materials}} \textbf{28}, 959--966 (2015).

\bibitem{eginligil2015dichroic}
M.~Eginligil, B.~Cao, Z.~Wang, X.~Shen, C.~Cong, J.~Shang, C.~Soci, and T.~Yu,
  \enquote{Dichroic spin{\textendash}valley photocurrent in monolayer
  molybdenum disulphide,} {\protect\JournalTitle{Nature Communications}}
  \textbf{6} (2015).

\bibitem{manzeli20172dtransition}
S.~Manzeli, D.~Ovchinnikov, D.~Pasquier, O.~V. Yazyev, and A.~Kis, \enquote{2d
  transition metal dichalcogenides,} {\protect\JournalTitle{Nature Reviews
  Materials}} \textbf{2} (2017).

\bibitem{li2019valley}
H.-K. Li, K.~Y. Fong, H.~Zhu, Q.~Li, S.~Wang, S.~Yang, Y.~Wang, and X.~Zhang,
  \enquote{Valley optomechanics in a monolayer semiconductor,}
  {\protect\JournalTitle{Nature Photonics}} \textbf{13}, 397--401 (2019).

\bibitem{li2020roomtemperature}
L.~Li, L.~Shao, X.~Liu, A.~Gao, H.~Wang, B.~Zheng, G.~Hou, K.~Shehzad, L.~Yu,
  F.~Miao, Y.~Shi, Y.~Xu, and X.~Wang, \enquote{Room-temperature valleytronic
  transistor,} {\protect\JournalTitle{Nature Nanotechnology}} \textbf{15},
  743--749 (2020).

\bibitem{mak2010atomically}
K.~F. Mak, C.~Lee, J.~Hone, J.~Shan, and T.~F. Heinz, \enquote{Atomically thin
  {MoS}$2$: A new direct-gap semiconductor,} {\protect\JournalTitle{Physical
  Review Letters}} \textbf{105} (2010).

\bibitem{bie2017mote2based}
Y.-Q. Bie, G.~Grosso, M.~Heuck, M.~M. Furchi, Y.~Cao, J.~Zheng, D.~Bunandar,
  E.~Navarro-Moratalla, L.~Zhou, D.~K. Efetov, T.~Taniguchi, K.~Watanabe,
  J.~Kong, D.~Englund, and P.~Jarillo-Herrero, \enquote{A {MoTe}$2$-based
  light-emitting diode and photodetector for silicon photonic integrated
  circuits,} {\protect\JournalTitle{Nature Nanotechnology}} \textbf{12},
  1124--1129 (2017).

\bibitem{mak2014thevalley}
K.~F. Mak, K.~L. McGill, J.~Park, and P.~L. McEuen, \enquote{The valley hall
  effect in {MoS}$2$ transistors,} {\protect\JournalTitle{Science}}
  \textbf{344}, 1489--1492 (2014).

\bibitem{yang2016electrically}
W.~Yang, J.~Shang, J.~Wang, X.~Shen, B.~Cao, N.~Peimyoo, C.~Zou, Y.~Chen,
  Y.~Wang, C.~Cong, W.~Huang, and T.~Yu, \enquote{Electrically tunable
  valley-light emitting diode ({vLED}) based on {CVD}-grown monolayer {WS}$2$,}
  {\protect\JournalTitle{Nano Letters}} \textbf{16}, 1560--1567 (2016).

\bibitem{wan2017epitaxial}
Y.~Wan, J.~Xiao, J.~Li, X.~Fang, K.~Zhang, L.~Fu, P.~Li, Z.~Song, H.~Zhang,
  Y.~Wang, M.~Zhao, J.~Lu, N.~Tang, G.~Ran, X.~Zhang, Y.~Ye, and L.~Dai,
  \enquote{Epitaxial single-layer {MoS}$2$ on {GaN} with enhanced valley
  helicity,} {\protect\JournalTitle{Advanced Materials}} \textbf{30}, 1703888
  (2017).

\bibitem{bernardi2013extraordinary}
M.~Bernardi, M.~Palummo, and J.~C. Grossman, \enquote{Extraordinary sunlight
  absorption and one nanometer thick photovoltaics using two-dimensional
  monolayer materials,} {\protect\JournalTitle{Nano Letters}} \textbf{13},
  3664--3670 (2013).

\bibitem{lee2018twodimensional}
E.~Lee, Y.~S. Yoon, and D.-J. Kim, \enquote{Two-dimensional transition metal
  dichalcogenides and metal oxide hybrids for gas sensing,}
  {\protect\JournalTitle{{ACS} Sensors}} \textbf{3}, 2045--2060 (2018).

\bibitem{schaibley2016valleytronics}
J.~R. Schaibley, H.~Yu, G.~Clark, P.~Rivera, J.~S. Ross, K.~L. Seyler, W.~Yao,
  and X.~Xu, \enquote{Valleytronics in 2d materials,}
  {\protect\JournalTitle{Nature Reviews Materials}} \textbf{1} (2016).

\bibitem{mccreary2017understanding}
K.~M. McCreary, M.~Currie, A.~T. Hanbicki, H.-J. Chuang, and B.~T. Jonker,
  \enquote{Understanding variations in circularly polarized photoluminescence
  in monolayer transition metal dichalcogenides,} {\protect\JournalTitle{{ACS}
  Nano}} \textbf{11}, 7988--7994 (2017).

\bibitem{gong2018nanoscale}
S.-H. Gong, F.~Alpeggiani, B.~Sciacca, E.~C. Garnett, and L.~Kuipers,
  \enquote{Nanoscale chiral valley-photon interface through optical spin-orbit
  coupling,} {\protect\JournalTitle{Science}} \textbf{359}, 443--447 (2018).

\bibitem{chen2016areview}
H.-T. Chen, A.~J. Taylor, and N.~Yu, \enquote{A review of metasurfaces: physics
  and applications,} {\protect\JournalTitle{Reports on Progress in Physics}}
  \textbf{79}, 076401 (2016).

\bibitem{rodrigo2016extraordinary}
S.~G. Rodrigo, F.~de~Le{\'{o}}n-P{\'{e}}rez, and L.~Mart{\'{i}}n-Moreno,
  \enquote{Extraordinary optical transmission: Fundamentals and applications,}
  {\protect\JournalTitle{Proceedings of the {IEEE}}} \textbf{104}, 2288--2306
  (2016).

\bibitem{genevet2017recent}
P.~Genevet, F.~Capasso, F.~Aieta, M.~Khorasaninejad, and R.~Devlin,
  \enquote{Recent advances in planar optics: from plasmonic to dielectric
  metasurfaces,} {\protect\JournalTitle{Optica}} \textbf{4}, 139 (2017).

\bibitem{beruete2019terahertz}
M.~Beruete and I.~J{\'{a}}uregui-L{\'{o}}pez, \enquote{Terahertz sensing based
  on metasurfaces,} {\protect\JournalTitle{Advanced Optical Materials}}
  \textbf{8}, 1900721 (2019).

\bibitem{watts2014terahertz}
C.~M. Watts, D.~Shrekenhamer, J.~Montoya, G.~Lipworth, J.~Hunt, T.~Sleasman,
  S.~Krishna, D.~R. Smith, and W.~J. Padilla, \enquote{Terahertz compressive
  imaging with metamaterial spatial light modulators,}
  {\protect\JournalTitle{Nature Photonics}} \textbf{8}, 605--609 (2014).

\bibitem{walter2017ultrathin}
F.~Walter, G.~Li, C.~Meier, S.~Zhang, and T.~Zentgraf, \enquote{Ultrathin
  nonlinear metasurface for optical image encoding,}
  {\protect\JournalTitle{Nano Letters}} \textbf{17}, 3171--3175 (2017).

\bibitem{bomzon2002spacevariant}
Z.~Bomzon, G.~Biener, V.~Kleiner, and E.~Hasman, \enquote{Space-variant
  pancharatnam{\textendash}berry phase optical elements with computer-generated
  subwavelength gratings,} {\protect\JournalTitle{Optics Letters}} \textbf{27},
  1141 (2002).

\bibitem{zhao2011manipulating}
Y.~Zhao and A.~Al{\`{u}}, \enquote{Manipulating light polarization with
  ultrathin plasmonic metasurfaces,} {\protect\JournalTitle{Physical Review B}}
  \textbf{84} (2011).

\bibitem{yu2012abroadband}
N.~Yu, F.~Aieta, P.~Genevet, M.~A. Kats, Z.~Gaburro, and F.~Capasso, \enquote{A
  broadband, background-free quarter-wave plate based on plasmonic
  metasurfaces,} {\protect\JournalTitle{Nano Letters}} \textbf{12}, 6328--6333
  (2012).

\bibitem{shitrit2013spinoptical}
N.~Shitrit, I.~Yulevich, E.~Maguid, D.~Ozeri, D.~Veksler, V.~Kleiner, and
  E.~Hasman, \enquote{Spin-optical metamaterial route to spin-controlled
  photonics,} {\protect\JournalTitle{Science}} \textbf{340}, 724--726 (2013).

\bibitem{langguth2015plasmonic}
L.~Langguth, A.~H. Schokker, K.~Guo, and A.~F. Koenderink, \enquote{Plasmonic
  phase-gradient metasurface for spontaneous emission control,}
  {\protect\JournalTitle{Physical Review B}} \textbf{92} (2015).

\bibitem{cotrufo2016spindependent}
M.~Cotrufo, C.~I. Osorio, and A.~F. Koenderink, \enquote{Spin-dependent
  emission from arrays of planar chiral nanoantennas due to lattice and
  localized plasmon resonances,} {\protect\JournalTitle{{ACS} Nano}}
  \textbf{10}, 3389--3397 (2016).

\bibitem{yan2017twisting}
C.~Yan, X.~Wang, T.~V. Raziman, and O.~J.~F. Martin, \enquote{Twisting
  fluorescence through extrinsic chiral antennas,} {\protect\JournalTitle{Nano
  Letters}} \textbf{17}, 2265--2272 (2017).

\bibitem{bliokh2015spinorbit}
K.~Y. Bliokh, F.~J. Rodr{\'{\i}}guez-Fortu{\~{n}}o, F.~Nori, and A.~V. Zayats,
  \enquote{Spin{\textendash}orbit interactions of light,}
  {\protect\JournalTitle{Nature Photonics}} \textbf{9}, 796--808 (2015).

\bibitem{chervy2018room}
T.~Chervy, S.~Azzini, E.~Lorchat, S.~Wang, Y.~Gorodetski, J.~A. Hutchison,
  S.~Berciaud, T.~W. Ebbesen, and C.~Genet, \enquote{Room temperature chiral
  coupling of valley excitons with spin-momentum locked surface plasmons,}
  {\protect\JournalTitle{{ACS} Photonics}} \textbf{5}, 1281--1287 (2018).

\bibitem{li2018tailoring}
Z.~Li, C.~Liu, X.~Rong, Y.~Luo, H.~Cheng, L.~Zheng, F.~Lin, B.~Shen, Y.~Gong,
  S.~Zhang, and Z.~Fang, \enquote{Tailoring {MoS}$2$ valley-polarized
  photoluminescence with super chiral near-field,}
  {\protect\JournalTitle{Advanced Materials}} \textbf{30}, 1801908 (2018).

\bibitem{jha2018spontaneous}
P.~K. Jha, N.~Shitrit, X.~Ren, Y.~Wang, and X.~Zhang, \enquote{Spontaneous
  exciton valley coherence in transition metal dichalcogenide monolayers
  interfaced with an anisotropic metasurface,} {\protect\JournalTitle{Physical
  Review Letters}} \textbf{121} (2018).

\bibitem{guddala2019valley}
S.~Guddala, R.~Bushati, M.~Li, A.~B. Khanikaev, and V.~M. Menon,
  \enquote{Valley selective optical control of excitons in 2d semiconductors
  using a chiral metasurface,} {\protect\JournalTitle{Optical Materials
  Express}} \textbf{9}, 536 (2019).

\bibitem{sun2019separation}
L.~Sun, C.-Y. Wang, A.~Krasnok, J.~Choi, J.~Shi, J.~S. Gomez-Diaz, A.~Zepeda,
  S.~Gwo, C.-K. Shih, A.~Al{\`{u}}, and X.~Li, \enquote{Separation of valley
  excitons in a {MoS}2 monolayer using a subwavelength asymmetric groove
  array,} {\protect\JournalTitle{Nature Photonics}} \textbf{13}, 180--184
  (2019).

\bibitem{rong2020photonic}
K.~Rong, B.~Wang, A.~Reuven, E.~Maguid, B.~Cohn, V.~Kleiner, S.~Katznelson,
  E.~Koren, and E.~Hasman, \enquote{Photonic rashba effect from quantum
  emitters mediated by a berry-phase defective photonic crystal,}
  {\protect\JournalTitle{Nature Nanotechnology}} \textbf{15}, 927--933 (2020).

\bibitem{kim2020asingle}
S.~Kim, Y.-C. Lim, R.~M. Kim, J.~E. Fr\"{o}ch, T.~N. Tran, K.~T. Nam, and
  I.~Aharonovich, \enquote{A single chiral nanoparticle induced valley
  polarization enhancement,} {\protect\JournalTitle{Small}} \textbf{16},
  2003005 (2020).

\bibitem{cao2012valleyselective}
T.~Cao, G.~Wang, W.~Han, H.~Ye, C.~Zhu, J.~Shi, Q.~Niu, P.~Tan, E.~Wang,
  B.~Liu, and J.~Feng, \enquote{Valley-selective circular dichroism of
  monolayer molybdenum disulphide,} {\protect\JournalTitle{Nature
  Communications}} \textbf{3} (2012).

\bibitem{lin2021electrically}
W.-H. Lin, P.~C. Wu, H.~Akbari, G.~R. Rossman, N.-C. Yeh, and H.~A. Atwater,
  \enquote{Electrically tunable and dramatically enhanced valley-polarized
  emission of monolayer {WS}$2$ at room temperature with plasmonic archimedes
  spiral nanostructures,} {\protect\JournalTitle{Advanced Materials}}
  \textbf{34}, 2104863 (2021).

\bibitem{guddala2021optical}
S.~Guddala, M.~Khatoniar, N.~Yama, W.~Liu, G.~S. Agarwal, and V.~M. Menon,
  \enquote{Optical analog of valley hall effect of 2d excitons in hyperbolic
  metamaterial,} {\protect\JournalTitle{Optica}} \textbf{8}, 50 (2021).

\bibitem{li2016tailoring}
Z.~Li, Y.~Li, T.~Han, X.~Wang, Y.~Yu, B.~Tay, Z.~Liu, and Z.~Fang,
  \enquote{Tailoring {MoS}$2$ exciton-plasmon interaction by optical spin-orbit
  coupling,} {\protect\JournalTitle{{ACS} Nano}} \textbf{11}, 1165--1171
  (2016).

\bibitem{guddala2021alloptical}
S.~Guddala, Y.~Kawaguchi, F.~Komissarenko, S.~Kiriushechkina, A.~Vakulenko,
  K.~Chen, A.~Al{\`{u}}, V.~M. Menon, and A.~B. Khanikaev, \enquote{All-optical
  nonreciprocity due to valley polarization pumping in transition metal
  dichalcogenides,} {\protect\JournalTitle{Nature Communications}} \textbf{12}
  (2021).

\bibitem{loren2023microscopic}
F.~Lor{\'{e}}n, G.~L. Paravicini-Bagliani, S.~Saha, J.~Gautier, M.~Li,
  C.~Genet, and L.~Mart{\'{\i}}n-Moreno, \enquote{Microscopic analysis of
  spin-momentum locking on a geometric phase metasurface,}
  {\protect\JournalTitle{Physical Review B}} \textbf{107} (2023).

\bibitem{lmm2008minimal}
L.~Mart{\'{\i}}n-Moreno and F.~J. Garc{\'{\i}}a-Vidal, \enquote{Minimal model
  for optical transmission through holey metal films,}
  {\protect\JournalTitle{Journal of Physics: Condensed Matter}} \textbf{20},
  304214 (2008).

\bibitem{deleon2008theory}
F.~de~Le{\'{o}}n-P{\'{e}}rez, G.~Brucoli, F.~J. Garc{\'{\i}}a-Vidal, and
  L.~Mart{\'{\i}}n-Moreno, \enquote{Theory on the scattering of light and
  surface plasmon polaritons by arrays of holes and dimples in a metal film,}
  {\protect\JournalTitle{New Journal of Physics}} \textbf{10}, 105017 (2008).

\bibitem{garciavidal2010light}
F.~J. Garcia-Vidal, L.~Martin-Moreno, T.~W. Ebbesen, and L.~Kuipers,
  \enquote{Light passing through subwavelength apertures,}
  {\protect\JournalTitle{Reviews of Modern Physics}} \textbf{82}, 729--787
  (2010).

\bibitem{fox2022generalized}
M.~Fox and Y.~Gorodetski, \enquote{Generalized approach to plasmonic phase
  modulation in topological bi-gratings,} {\protect\JournalTitle{Applied
  Physics Letters}} \textbf{120}, 031105 (2022).

\bibitem{vial2005improved}
A.~Vial, A.-S. Grimault, D.~Mac\'{\i}as, D.~Barchiesi, and M.~L. de~la
  Chapelle, \enquote{Improved analytical fit of gold dispersion: Application to
  the modeling of extinction spectra with a finite-difference time-domain
  method,} {\protect\JournalTitle{Phys. Rev. B}} \textbf{71}, 085416 (2005).

\bibitem{liu2014optical}
H.-L. Liu, C.-C. Shen, S.-H. Su, C.-L. Hsu, M.-Y. Li, and L.-J. Li,
  \enquote{Optical properties of monolayer transition metal dichalcogenides
  probed by spectroscopic ellipsometry,} {\protect\JournalTitle{Applied Physics
  Letters}} \textbf{105}, 201905 (2014).

\bibitem{lorchat2018roomtemperature}
E.~Lorchat, S.~Azzini, T.~Chervy, T.~Taniguchi, K.~Watanabe, T.~W. Ebbesen,
  C.~Genet, and S.~Berciaud, \enquote{Room-temperature valley polarization and
  coherence in transition metal dichalcogenide{\textendash}graphene van der
  waals heterostructures,} {\protect\JournalTitle{{ACS} Photonics}} \textbf{5},
  5047--5054 (2018).

\bibitem{lorchat2020filtering}
E.~Lorchat, L.~E.~P. L{\'{o}}pez, C.~Robert, D.~Lagarde, G.~Froehlicher,
  T.~Taniguchi, K.~Watanabe, X.~Marie, and S.~Berciaud, \enquote{Filtering the
  photoluminescence spectra of atomically thin semiconductors with graphene,}
  {\protect\JournalTitle{Nature Nanotechnology}} \textbf{15}, 283--288 (2020).

\bibitem{chen2017valleypolarized}
Y.-J. Chen, J.~D. Cain, T.~K. Stanev, V.~P. Dravid, and N.~P. Stern,
  \enquote{Valley-polarized exciton{\textendash}polaritons in a monolayer
  semiconductor,} {\protect\JournalTitle{Nature Photonics}} \textbf{11},
  431--435 (2017).

\bibitem{dufferwiel2018valley}
S.~Dufferwiel, T.~P. Lyons, D.~D. Solnyshkov, A.~A.~P. Trichet, A.~Catanzaro,
  F.~Withers, G.~Malpuech, J.~M. Smith, K.~S. Novoselov, M.~S. Skolnick, D.~N.
  Krizhanovskii, and A.~I. Tartakovskii, \enquote{Valley coherent
  exciton-polaritons in a monolayer semiconductor,}
  {\protect\JournalTitle{Nature Communications}} \textbf{9} (2018).

\bibitem{schneider2018twodimensional}
C.~Schneider, M.~M. Glazov, T.~Korn, S.~H\"{o}fling, and B.~Urbaszek,
  \enquote{Two-dimensional semiconductors in the regime of strong light-matter
  coupling,} {\protect\JournalTitle{Nature Communications}} \textbf{9} (2018).

\bibitem{sun2020selectively}
J.~Sun, H.~Hu, D.~Pan, S.~Zhang, and H.~Xu, \enquote{Selectively depopulating
  valley-polarized excitons in monolayer {MoS}$2$ by local chirality in single
  plasmonic nanocavity,} {\protect\JournalTitle{Nano Letters}} \textbf{20},
  4953--4959 (2020).

\bibitem{zheng2023electroninduced}
L.~Zheng, Z.~Dang, D.~Ding, Z.~Liu, Y.~Dai, J.~Lu, and Z.~Fang,
  \enquote{Electron-induced chirality-selective routing of valley photons via
  metallic nanostructure,} {\protect\JournalTitle{Advanced Materials}} p.
  2204908 (2023).

\bibitem{gao2023atomicallythin}
T.~Gao, M.~von Helversen, C.~Ant{\'{o}}n-Solanas, C.~Schneider, and T.~Heindel,
  \enquote{Atomically-thin single-photon sources for quantum communication,}
  {\protect\JournalTitle{npj 2D Materials and Applications}} \textbf{7} (2023).

\end{thebibliography}

\end{document}